\documentclass[proof]{pasj00}
\draft

\newcommand{\axp}{1E~1547.0$-$5408}
\newcommand{\ergs}{erg\,s$^{-1}$}
\newcommand{\ergcm}{erg\,cm$^{-2}$}
\newcommand{\ergcms}{erg\,cm$^{-2}$\,s$^{-1}$}
\usepackage{lineno}

\begin{document}

\SetRunningHead{Author(s) in page-head}{Running Head}
\Received{$\langle$reception date$\rangle$}
\Accepted{$\langle$acception date$\rangle$}
\Published{$\langle$publication date$\rangle$}

\title{Sub-MeV Band Observation of a Hard Burst from AXP~\axp\ with the Suzaku Wide-band All-sky Monitor}

\author{Tetsuya~Yasuda\altaffilmark{1},
  Wataru~B.~Iwakiri\altaffilmark{2},
  Makoto~S.~Tashiro\altaffilmark{1},
  Yukikatsu~Terada\altaffilmark{1},
  Tomomi~Kouzu\altaffilmark{1},
  Teruaki~Enoto\altaffilmark{2}$^{,}$\altaffilmark{3},
  Yujin~E.~Nakagawa\altaffilmark{4},
  Aya~Bamba\altaffilmark{5},
  Yuji~Urata\altaffilmark{6},
  Kazutaka~Yamaoka\altaffilmark{7}$^{,}$\altaffilmark{8},
  Masanori~Ohno\altaffilmark{9},
  Sinpei~Shibata\altaffilmark{10},
  Kazuo~Makishima\altaffilmark{11},
  and~The~Suzaku-WAM~team
}
\altaffiltext{1}{Graduate School of Science and Engineering, Saitama University, 255 Shimo-Okubo, Sakawa, Saitama, Saitama 338-8570, Japan}
\altaffiltext{2}{High Energy Astrophysics Laboratory, the Institute of Physical and Chemical Research, 2-1 Hirosawa, Wako, Saitama 351-0198, Japan}
\altaffiltext{3}{NASA Goddard Space Flight Center, Greenbelt, MD 20771, USA}
\altaffiltext{4}{Institute of Space and Astronautical Science (ISAS), Japan Aerospace Exploration Agency (JAXA), 3-1-1 Yoshinodai, Chuo, Sagamihara, Kanagawa 252-5210, Japan}
\altaffiltext{5}{Department of Physics and Mathematics, Aoyama Gakuin University 5-10-1 Fuchinobe Chuo-ku, Sagamihara, Kanagawa 252-5258, Japan}
\altaffiltext{6}{Institute of Astronomy, National Central University, Chung-Li 32054, Taiwan}
\altaffiltext{7}{Solar-Terrestrial Environment Laboratory, Nagoya University, Furo-cho, Chikusa-ku, Nagoya, Aichi 464-8601, Japan}
\altaffiltext{8}{Division of Particle and Astrophysical Science, Graduate School of Science, Nagoya University, Furo-cho, Chikusa-ku, Nagoya, Aichi 464-8601, Japan}
\altaffiltext{9}{Department of Physical Science, Hiroshima University, 1-3-1 Kagamiyama, Higashi-Hiroshima, Hiroshima 739-8526, Japan}
\altaffiltext{10}{Department of Physics, Yamagata University, Kojirakawa, Yamagata 990-8560, Japan}
\altaffiltext{11}{Department of Physics, The University of Tokyo, 7-3-1 Hongo, Bunkyo, Tokyo 113-0033, Japan}
\email{yasuda@heal.phy.saitama-u.ac.jp}

\maketitle

\begin{abstract}
The 2.1-s anomalous X-ray pulsar 1E~1547.0$-$5408 exhibited an X-ray outburst on 2009 January 22, emitting a large number of short bursts.
The wide-band all-sky monitor (WAM) on-board Suzaku detected at least 254 bursts in the 0.16--6.2\,MeV band over the period of January 22 00:57--17:02 UTC from the direction of \axp.
One of these bursts, which occurred at 06:45:13, produced the brightest fluence in the 0.5--6.2\,MeV range, with an averaged 0.16--6.2\,MeV flux and extrapolated 25\,keV--2\,MeV fluence of about $1 \times 10^{-5}$\,\ergcms\ and about $3 \times 10^{-4}$\,\ergcm, respectively.
After pile-up corrections, the time-resolved WAM spectra of this burst were well-fitted in the 0.16--6.2\,MeV range by two-component models; specifically, a blackbody plus an optically thin thermal bremsstrahlung or a combination of a blackbody and a power-law component with an exponential cutoff.
These results are compared with previous works reporting the persistent emission and weaker short bursts followed by the same outburst.
\end{abstract}



\section{Introduction}
\noindent
Magnetars, which are observed as soft gamma repeaters (SGRs) and anomalous X-ray pulsars (AXPs; for reviews, see \cite{woods2006,kaspi2007,mereghetti2008}), are considered to be isolated neutron stars with strong magnetic fields of 10$^{13}$--10$^{15}$\,Gauss, which exceeds the quantum critical field $B_{\rm Q}$ of about $4.4 \times $ 10$^{13}$\,Gauss \citep{duncan1992,thompson1995}.
They are known to exhibit emissions as sporadic bursts, which are classified into three kinds according to their luminosities and durations: ``giant flares'', ``intermediate flares,'' and ``short bursts.''

Spectra of all three giant flares observed to date are characterized by optically thin thermal bremsstrahlung (OTTB) with plasma temperature $kT$ from a few tens to hundreds of keV \citep{hurley1999,palmer2005}.
However, some observational results in the energy spectra have indicated different spectral shapes from OTTB, such as a single hard power-law model with an exponential cutoff function (PLE, \cite{frederiks2007}), and a hard power-law extending to above 1 MeV in energy range \citep{boggs2007,frederiks2007}.
X-ray spectra of intermediate flares and short bursts are usually reproduced by a combination of two-blackbody components (2BB) with temperatures of $kT_{\rm low} \sim 2$--$4$\,keV and $kT_{\rm high} \sim 8$--$15$\,keV \citep{feroci2004,olive2004,nakagawa2007,esposito2007,mereghetti2008}.
\citet{nakagawa2011} and \citet{enoto2012} discovered high-energy photons extending into the sub-MeV band in spectra of accumulated weak short bursts from SGR~0501$+$4516 and AXP~\axp, respectively.
Since the spectral shapes of the bursts in the sub-MeV band are complicated and not yet clarified, the spectra need to be better quantified in order to investigate the emission mechanisms.

AXP~\axp\ (SGR~J1550$-$5418, PSR~J1550$-$5418), presented in this paper, is a magnetar associated with a young supernova remnant G327.24$-$0.13 \citep{gelfand2007}.
According to the spin period ($\sim 2.1$\,s) and spin-down rate ($\sim$ 4.8 $\times 10^{-11}$ s/s) reported by \citet{dib2012}, the dipole surface magnetic field strength and characteristic age are estimated to be about 3.2 $\times $10$^{14}$ Gauss and $ 0.69$\,kyr, respectively.
These features make this object relatively young and the fastest-known rotating magnetar\footnote{http://www.physics.mcgill.ca/\~{}pulsar/magnetar/main.html}.
On 2009 January 22, the object entered an active phase and produced a large number of short bursts, as detected by Swift \citep{gronwall2009}, Fermi \citep{connaughton2009}, INTEGRAL \citep{savchenko2009,mereghetti2009b}, Konus-Wind (\cite{golenetskii2009a}, 2009b, 2009c), RHESSI \citep{bellm2009}, and Suzaku \citep{terada2009}.
This activity was observed by several high-energy missions, creating a good opportunity for investigating the broadband spectra of magnetar short bursts and intermediate flares in detail.
Broadband spectral properties have been reported by several authors (e.g., \cite{horst2012,lin2012,younes2014}), but these observations are limited to energies below about $200$ keV.

In this paper, we focus on spectral analysis of an energetic burst that occurred at UTC 2009 January 22 06:45:13 and was observed by the Suzaku wide-band all-sky monitor (WAM; \cite{yamaoka2009}).
The large effective area and wide energy range available from the WAM enable us to investigate the 0.16--6.2 MeV spectra of this energetic event.
We also compare our results with spectra having persistent emission and stacked weak short bursts, as observed from this source by Suzaku (\cite{enoto2010a}, 2012).
In the following sections, we assume a distance to \axp\ of $d \sim 4$ kpc \citep{tiengo2010} and use  \texttt{hxdbstjudge}, \texttt{hxdmkwamlc}, and \texttt{hxdmkwamspec}, which are standard FTOOLS included in the HEADAS software package version 6.13.
The quoted errors are for a $90\%$ confidence level.

\section{Observation}
\noindent
The WAM is an active shield crystal in the hard X-ray detector (HXD, \cite{takahashi2007,kokubun2007}) on-board the Suzaku satellite \citep{mitsuda2007}.
It comprises four walls (WAM-0, WAM-1, WAM-2 and WAM-3) made of bismuth orthogermanate Bi$_{4}$Ge$_{3}$O$_{12}$ (BGO) crystals, which were initially designed to measure the sub-MeV gamma-ray spectrum in the nominal range of 50\,keV to 5\,MeV. However, the energy range has shifted to range between 70 keV and 6.2 MeV range of photon energy due to long term degradation of gain at the time of measurement.
The four detectors form the lateral sides of a square tube, and each has an acceptance of 2$\pi$ sr with a nominal direction having effective area of 400 cm$^{2}$ at 1 MeV.
Their nominal directions have azimuth angles of $\phi = 90^\circ$ (WAM-0), $\phi = 0^\circ$ (WAM-1), $\phi = 180^\circ$ (WAM-2), and $\phi = 270^\circ$ (WAM-3), all with a zenith angle at $\theta = 90^\circ$, where $\theta = 0^\circ$ is defined as the HXD on-axis.
Among currently working gamma-ray spectrometers on-board astronomical satellites, the WAM has the largest effective area\footnote{Among photon counting detectors, the INTEGRAL anti-coincident shield has the largest effective area \citep{mereghetti2005}} for energies within 0.3--6.2\,MeV.
Thanks to these characteristics, the WAM has so far detected a large number of gamma-ray bursts and solar flares (e.g., \cite{endo2010,tashiro2012,urata2012,urata2014}).

The WAM produces two types of datasets: burst (BST) data and transient (TRN) data.
The BST data are recorded in four energy channels with time resolution of 1/64\,s, but they cover only 64 s around when sudden changes in count rates trigger the BST data acquisition in orbit.
Only one set of BST data can be stored in the on-board buffer before the spacecraft data recorder reads it out during passage through the South Atlantic Anomaly (SAA).
After the readout, the onboard buffer becomes recordable again.
Because of this limitation, we can obtain at most about ten BST data sets in a day, but we lose later BST data if other flares occur soon after recording is triggered.
In contrast, the TRN data are continuously accumulated with a 1-s time resolution in 55 energy channels covering 70\,keV--6.2\,MeV, except for during bias-voltage reduction due to SAA passage.
The absolute time accuracies of these two datasets are $320~\mu$s \citep{terada2008}.

\begin{figure}
  \begin{center}
      \FigureFile(80mm,80mm){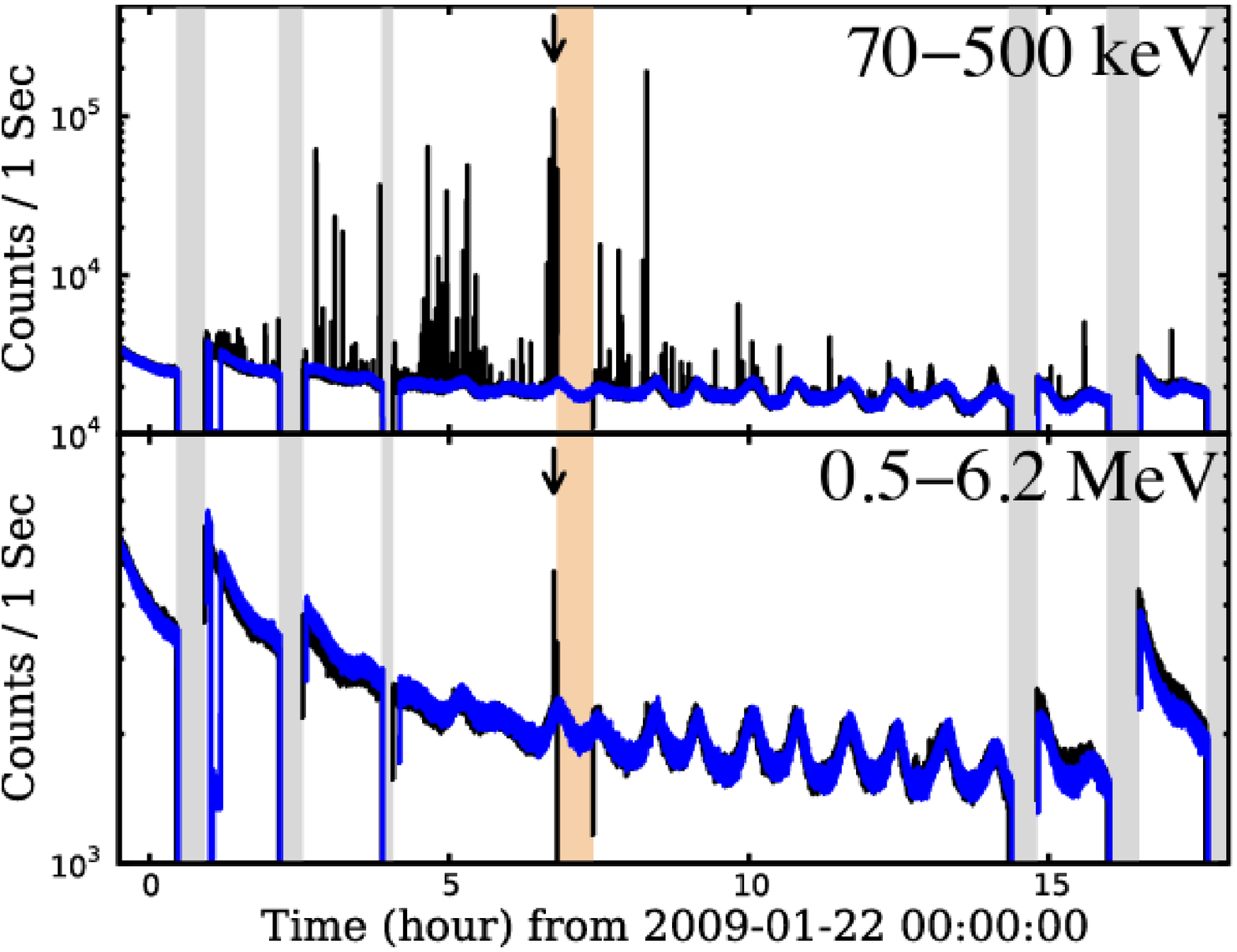}
      \FigureFile(80mm,80mm){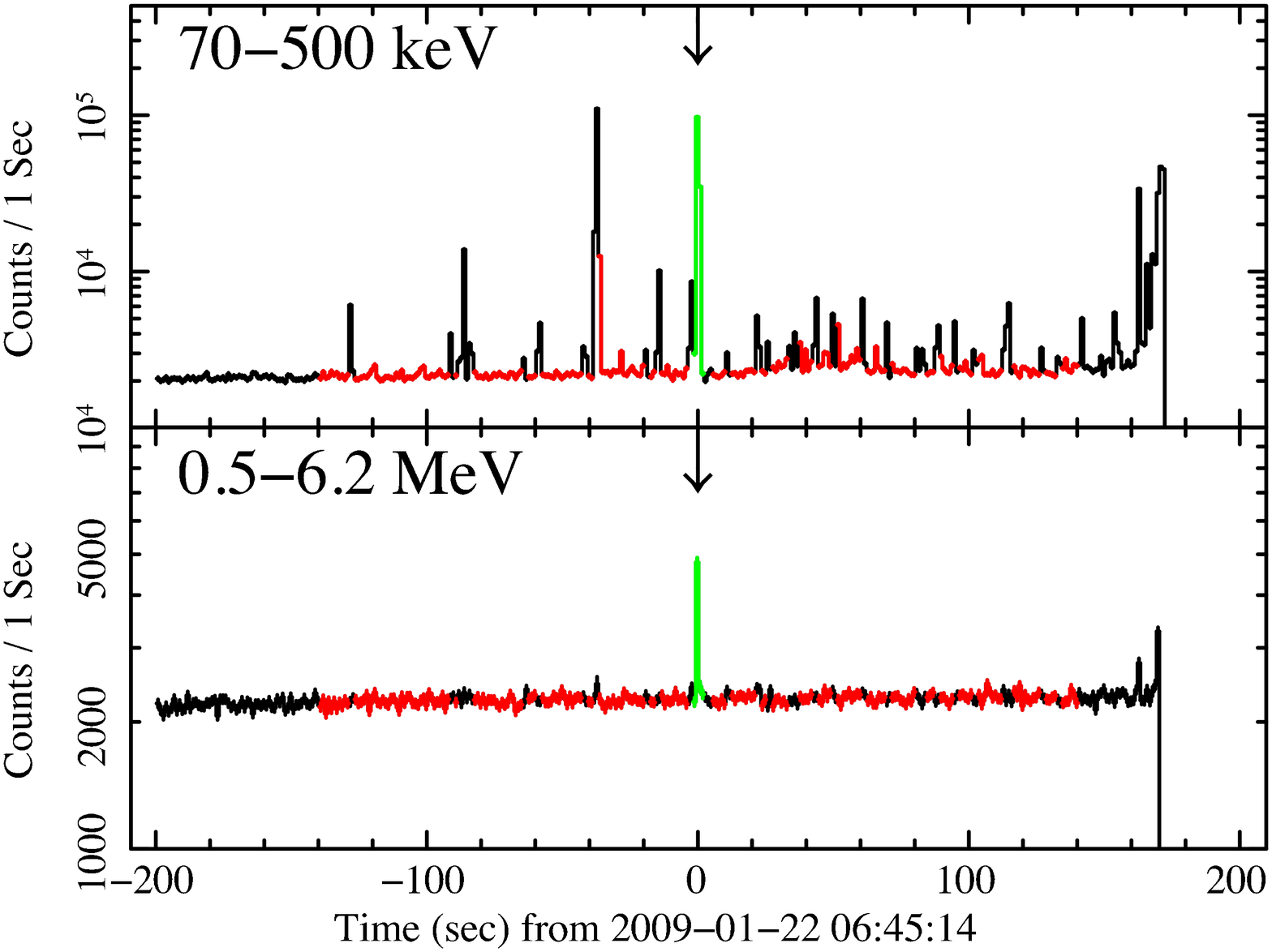}
  \end{center}
  \caption{Suzaku/WAM-0 light curves of two bands on 2009 January 22.
  Employed time binning is 1.0 s.
  Background components are not subtracted but dead time is corrected.
  For comparison blue lines show the light curves for the following day when there were no detected events.
  Left panels show the entire day while the right panels show the light curves near the hardest burst. 
  Arrows indicate the hardest burst which occurred at 06:45:14.
  Data gaps are due to passages through the SAA (gray zone) and the time while the safety function was active after being triggered by the bright event at 06:48:04 (orange zone).
  The WAM made no observations during these periods.
  To perform spectral analysis in \S\ref{sec:withPileup}, time regions of green and red lines are accumulated as source and background spectra, respectively.
  }
  \label{fig:lightCurve}
\end{figure}

On 2009 January 22, \citet{gronwall2009} first reported short burst activities from \axp.
At the same time, the WAM had successfully record a series of bright burst-like events.
Figure \ref{fig:lightCurve} shows the light curves of WAM-0 TRN data during this activity (observational ID = 703049010, covering January 19 23:30--22 22:32).
A long-term trend on the hour scale is normally seen on calm days which is caused by fluctuations in the non-X-ray background induced by trapped charge particles.
In addition to this background variation, we also see a large number of extremely bright bursts for about $16$ hours.
Although the WAM is a non-imaging detector, we consider these bursts to have come from \axp\ because many of them were simultaneously detected by several high-energy missions, including INTEGRAL, Swift, Fermi, and Konus/Wind.
A detailed examination of the origin is described in \S\ref{sec:detection} and \S\ref{sec:angle}.
Note that since the highest fluence burst among those detected by the INTEGRAL anti-coincident shield (ACS) with pulsating tails at 06:48:04 \citep{mereghetti2009a} was extremely bright, it triggered the WAM safety functionality to turn off the detector high voltage (orange zone in Figure \ref{fig:lightCurve}).

\section{Temporal Analysis}
\label{sec:analysis}

\subsection{Short Burst Detection}
\label{sec:detection}
In order to extract the short bursts from the WAM light curves, we used \texttt{hxdbstjudge} with input parameters of \texttt{bgd\_integ\_time = 8}, \texttt{delta\_t = 1} and \texttt{sigma = 5.5}, which produces detection criteria of (1) calculating the average count rate every 8\,s, which is treated as the background level before any given time, and (2) comparing it with the count rate every 1\,s, which could include both flare signal and background components, then (3) judging the burst when the flare signal (2) exceeds 5.5-fold the standard deviation of the background level (1).
Consequently, for the WAM light curves in Figure \ref{fig:lightCurve}, 254, 176, 39, and 41 events were successfully detected by WAM-0, WAM-1, WAM-2, and WAM-3, respectively, in energy channels 2--11 (70--500\,keV).
In the upper energy band in energy channels 12--54, which covers 0.5--6.2\,MeV, only three events were detected that satisfied the criteria above. These occurred at UTC January 22 06:45:13, 06:47:56, and 08:17:29.
The derived degrees of significance are 39.2, 6.1 and 8.5 sigma.
Among all detected bursts, 5 events, with trigger times of UTC 01:28:59, 02:46:56, 04:34:52, 15:10:34, and 17:02:55, were stored in the BST data format.
The $T_{90}$ durations, that is, the time to accumulate between 5$\%$ and 95$\%$ of the counts, of the events were distributed from 0.13 s to 2.0 s, and were reported in a gamma-ray burst (GRB) Coordinate Network circular \citep{terada2009}.
No bursts that satisfied the criteria were detected in the WAM-0 light curves on the day before or the day after the activity.

The time duration between the first detection at 00:57:20 and the last one at 17:02:56 is 58\,ks, while the total off-time due to SAA passage and the WAM safety functionality is 8\,ks.
If we assume that these bursts come from \axp, no occultation of the object by the earth is expected from the satellite attitude during the observations.
Therefore net exposure of the target source was 50\,ks, and WAM-0 detected the short bursts at a frequency of $5\times10^{-3}$\,s$^{-1}$ on average.
The ACS detected 233 bursts from UTC January 21 18:11 to January 23 04:27 \citep{mereghetti2009a}.

\subsection{Estimation of the Incident Angles}
\label{sec:angle}
\begin{figure}
  \begin{center}
    \begin{center}
      \FigureFile(80mm,80mm){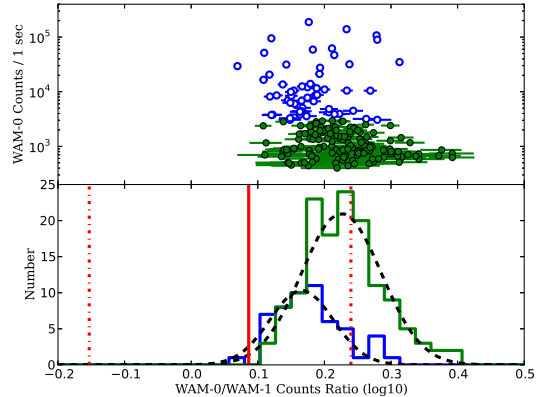}
    \end{center}
    \caption{Scatter plot between 1-s peak-count rate by WAM-0 in the 70--500\,keV band and the count rate by WAM-1 (top), and histogram of the count rate ratio (bottom).
    Blue circles (top) and histogram entries (bottom) are brighter events, and green ones are the rest.
    Black dashed lines show the best-fit lognormal Gaussian functions.
    Vertical red solid and dashed-dotted lines indicate the expected count rate ratio from \axp\ and the systematic error of 30\%, respectively.}
    \label{fig:hist}
  \end{center}
\end{figure}

Although the WAM has no imaging capability, the count rate ratio between WAM-0 and WAM-1 provides information about the angle of incidence of irradiating photons.
We examined the ratio of count rates between WAM-0 and WAM-1 for the bursts detected by both sides in energy channels 2--11 to check consistency with the assumption that they are from \axp.
The top panel of Figure \ref{fig:hist} shows a scatter plot of the count rate in WAM-0 and the ratio of count rate between WAM-0 and WAM1.
Since the statistical errors are comparatively large for dim bursts, distributions of the count rate ratios are shown separately for events brighter or weaker than $3.3 \times 10^{3}$\,counts\,s$^{-1}$ with WAM-0.
The results follow lognormal Gaussian shapes well, with mean values of $0.17 \pm 0.01$ and $0.23 \pm0.01$, and sigmas of $0.05 \pm 0.01$ and $0.06 \pm 0.01$ for brighter and weaker events, respectively.
The dispersion of the weaker burst distribution is explained by the statistical Poisson distribution law.
The difference in mean values between the two distributions is mainly due to a pile-up effect in the WAM analog signal processing unit.
For comparison, the expected ratio of count rates can be calculated from the assumed target position (i.e., \axp) and the energy response function of the WAM detectors.
According to the location of the object by \citet{deller2012} and the attitude of the satellite, the incident angles from the object to the WAM are zenith angle $\theta = 59.9^\circ$ and azimuthal angle $\phi = 51.4^\circ$.
This requires a count rate ratio of $0.09^{+0.15}_{-0.24}$ in a lognormal frame (shown in red in Figure \ref{fig:hist} bottom), with the errors, as estimated from the systematic error in the effective areas of the WAM response matrix, at 30\% \citep{yamaoka2009}.
As a result, the observed ratios of count rates are both consistent with the expected values under the assumption that the signals come from \axp\ (to within systematic error bounds), although the centroids are not well aligned.
We also checked the GRB Coordinate Network circular\footnote{http://gcn.gsfc.nasa.gov/gcn3\_archive.html} and archive of solar flares\footnote{http://www.lmsal.com/solarsoft/latest\_events\_archive.html}, and confirmed that there were no reports of other astronomical transient events in the same period.
We therefore concluded that all the bursts detected by the WAM came from the direction of the magnetar.
We hereafter use data from the two well-irradiated sides WAM-0 and WAM-1 in the analysis.

\subsection{Hardest Burst}
\label{sec:hardestburst}

Among all the bursts detected by WAM-0, only three events are also detected in the high-energy band between 0.5--6.2\,MeV, as described in \S\ref{sec:detection}.
Detection times were UTC January 22 06:45:13, 06:47:56, and 08:17:29.
The hardness ratios of these, defined as the ratio of the count rate in the 0.5--6.2\,MeV band to that in the 70--500\,keV band, are 0.021 $\pm 0.001$, 0.013 $\pm 0.001$, and 0.0026 $\pm 0.0001$, respectively.
In particular, the burst observed at 06:45:13 is the hardest (arrows in Figure \ref{fig:lightCurve}).
We hereafter call this event ``the hardest burst'' and present spectral analysis of this burst in the following sections.
The ratio of count rates between WAM-0 and WAM-1 is $0.12 \pm 0.01$ in a lognormal frame, which is within 1.1 $\sigma$ from the mean of the distribution of brighter bursts.

\begin{figure}
  \begin{center}
   \FigureFile(80mm,80mm){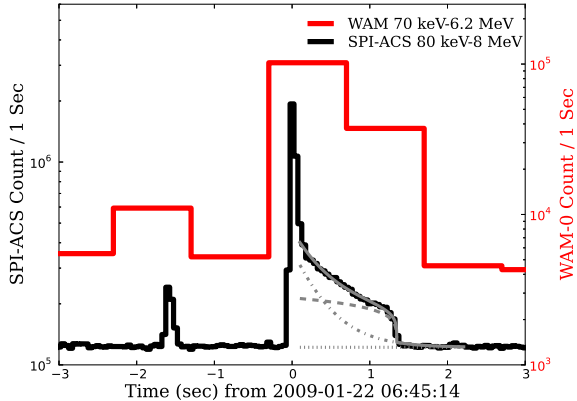}
  \end{center}
  \caption{Suzaku/WAM-0 (red solid line) and INTEGRAL/ACS (black solid line) light curves of the hardest burst with correction of the time lag of 0.410\,s between the two light curves.
	Gray lines represent the trapped fireball component (dashed line), an exponential function (dash-dotted line), the sum of both models (solid line), and background constant (dotted line). See the text (in discussion \S \ref{sec:withPileup}).
  	The background component is added to the first three lines.
	}
  \label{fig:detailedLightCurve}
\end{figure}

The hardest burst was observed only in the TRN data (i.e., no BST data are available) because the previous event had been stored and the hardest burst was not able to trigger the high-time-resolution BST mode.
\citet{mereghetti2009a} and \citet{savchenko2010} reported that INTEGRAL/ACS detected this burst at UTC 06:45:13.9 as the brightest event and labeled it with identifier number 121 and {\bfseries b} with properties of a duration of 1.45\,s, peak flux of $>26.3 \times 10^{-5}$\,\ergcms, and fluence of $> 4.6 \times 10^{-5}$\,\ergcm.
Those properties are comparable with the intermediate class among the three kinds of magnetar bursts.
The detailed ACS light curve (Trigger ID = 2009-01-22 06:44:36) is published in the Heavens archive\footnote{http://isdc.unige.ch/heavens/} with 50-ms time resolution in the 80\,keV--8\,MeV energy range.
Although the ACS has no energy information about each photon, it provides a high-time-resolution light curve in similar energy bands to those from WAM.
Figure \ref{fig:detailedLightCurve} shows the light curves obtained by WAM and ACS.
We observe an apparent time lag, which is considered to be due to time of flight of photons between Suzaku and INTEGRAL.
In the following analysis, we corrected for this time difference of 410\,ms calculated from the orbit information of the both satellites at the hardest burst.
According to \citet{yamaoka2009} using the BST data triggered by classical GRBs, a correction accuracy of the time of flight is $-2 \pm36$\,ms for bright events.

\section{Spectral Analysis}

\subsection{Data Selection for Spectral Analysis}
\label{sec:dataSelection}

\begin{figure}
  \begin{center}
   \FigureFile(90mm,90mm){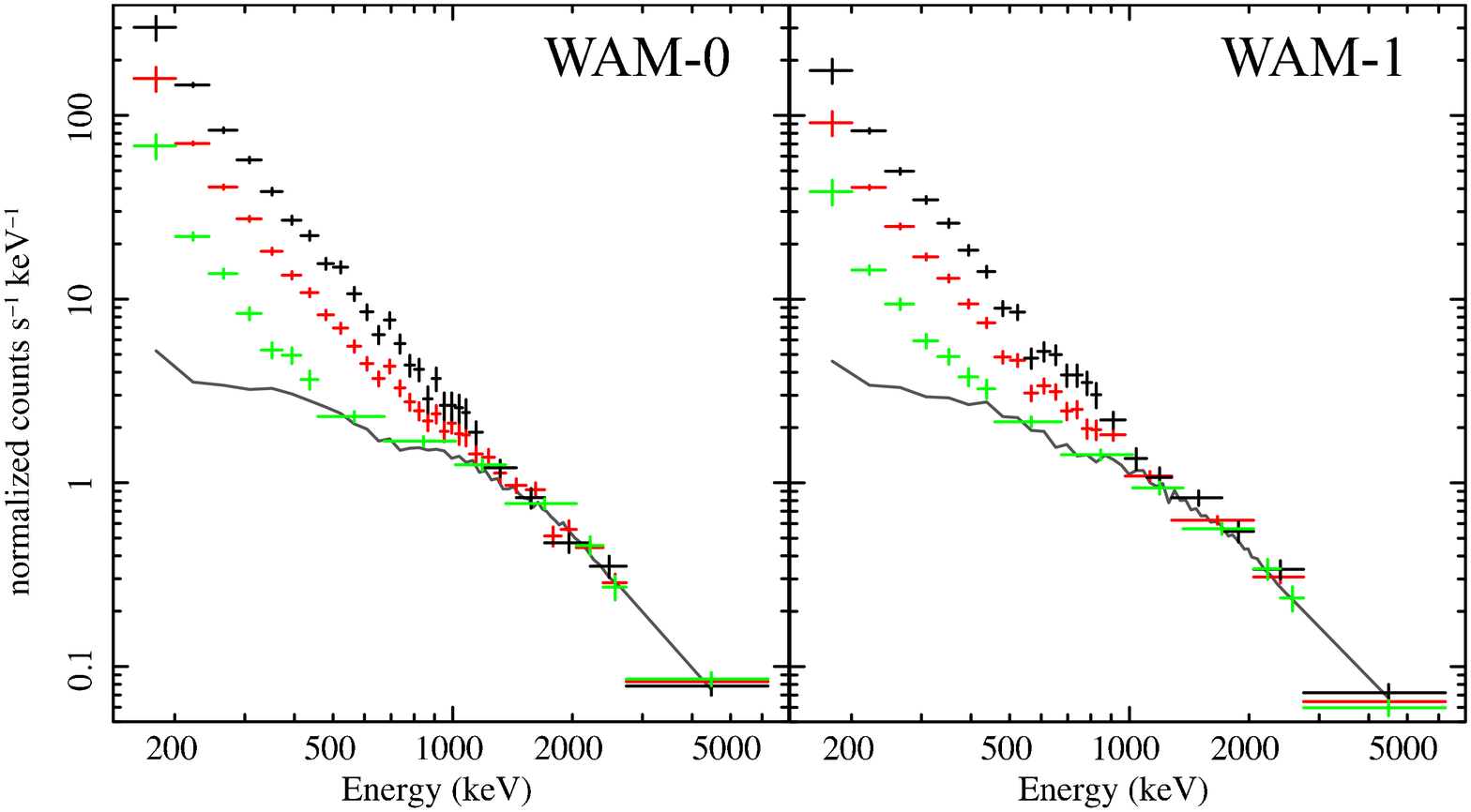}
  \end{center}
  \caption{Suzaku/WAM spectra of three time intervals, shown without removing the instrumental responses and subtracting background components.
  Black, red, green, and gray colors indicate peak, average, tail, and time-averaged background spectra, respectively.
	}
  \label{fig:rawSpec}
\end{figure}

To perform spectral analysis of the hardest burst defined in \S \ref{sec:hardestburst}, we extracted energy spectra of three source intervals and estimated the background spectra of WAM-0 and WAM-1 as follows.
We extracted time-averaged spectra of the hardest burst between 06:45:13.3 (Modified Julian Day = 54853.28140379, hereafter $T_{0}$) and 06:45:15.3 (54853.28142694) from TRN format data.
Since we used data with 1-s time resolution and the duration of the hardest burst is 1.5 s \citep{mereghetti2009a}, the first 1-s region ($T_{0}$--$T_{0}+1$\,s) is defined as the peak and the following region ($T_{0}+1$\,s--$T_{0}+2$\,s) as the tail.
Background spectra are extracted from the average of before and after the hardest burst time regions, specifically, from $T_{0}-140$ s to $T_{0}-4$ s and from $T_{0}+4$ s to $T_{0}+140$ s, respectively, avoiding detected short bursts with WAM-0 and WAM-1, shown in \S \ref{sec:detection}.
The time intervals used for the background are shown in Figure \ref{fig:lightCurve} as red lines.
Figure \ref{fig:rawSpec} shows the extracted spectra in count rate space.

\begin{figure}
  \begin{center}
   \FigureFile(80mm,80mm){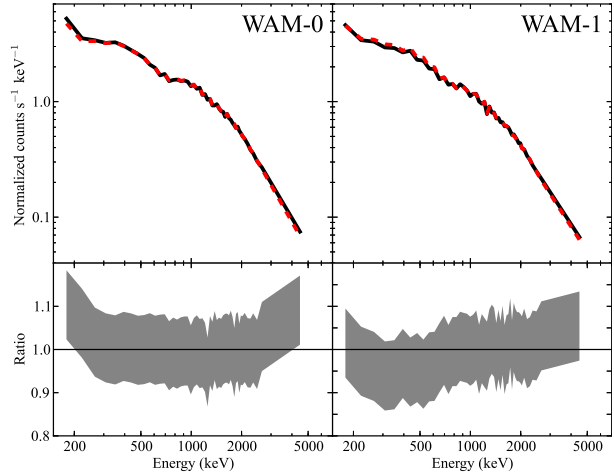}
  \end{center}
  \caption{Background spectra estimated from the time-averaged observations (black solid lines) and COR method (red dashed lines).
  The lower panels show the ratio between the former and latter spectra.
  The shaded area indicates the systematic error of 8\% in the COR background.
	}
  \label{fig:compareBgd}
\end{figure}

Since it is difficult to avoid contamination by weak undetected bursts using the above background estimation, we compared the obtained background spectra with those from another method, using cut-off rigidity (COR, \cite{endo2010}) to investigate whether the estimated time-averaged background spectra are suitable for performing spectral analysis.
In the COR method, the background flux and spectra are estimated from the data observed one day before and after the target event.
The reproducibility of this method is reported to be typically about 7--8\% \citep{endo2010}.
We therefore added 8\% systematic error to COR background spectra for comparison with the time-averaged background spectra.
Figure \ref{fig:compareBgd} shows the estimated background spectra.
The count rates of each energy channel in the background spectra are consistent within the error in the energy range above 200\,keV, while in the range below 200\,keV, the count rate of the time-averaged background is higher than the COR background by about $15$ percentage points.
This is thought to be due to the time-averaged background including weak unresolved bursts.
However since the response matrix uncertainty below 160\,keV is insufficient under current calibration \citep{yamaoka2009}, we cannot examine the possible unresolved short bursts in detail.
In order to perform spectral analysis, we therefore ignored the range below 160\,keV and added a systematic error of 15\% to energy bins in the range between 160 and 200\,keV, and applied the time-averaged background spectra to the three intervals.

\subsection{Response matrices and pile-up corrections}
\label{sec:respPileup}
The response matrices of the WAM depend on angle of incidence of photons since the WAM is mounted inside a spacecraft and the observed spectra are heavily affected by absorption by the satellite structure.
In order to calculate the responses of the detectors, we use a Geant4-based Monte-Carlo simulation code \citep{ohno2005,terada2005,ozaki2005}.
The uncertainties of the matrices have been studied before launch by ground calibration measurements of radio-isotope irradiation tests at various incident angles \citep{terada2005} and also confirmed by in-orbit cross calibration between the Swift-BAT, the Konus-Wind, and the Suzaku-WAM using GRB spectra \citep{sakamoto2011}, and by an Earth-occultation technique using Crab nebula spectra \citep{yamaoka2009}.
The absolute flux for the incident angle of \axp\ ($\theta = 59.9^\circ$, $\phi = 51.4^\circ$) has uncertainty of about $30$\%, which can be seen by referring to Figure 18 in \citet{yamaoka2009}.
Since the satellite structure on the WAM-0 side is not as complex as that on the other sides, the effective area for the WAM-0 is the most reliable among those of the sides.
Therefore, we leave the normalization factor of WAM-1 with respect to the WAM-0 side as a free parameter in the spectral fitting below.

The hardest burst was too bright to measure the exact peak flux by ACS observation \citep{mereghetti2009a}.
The obtained peak count rate of WAM-0 exceeded $1.6 \times 10^{5}$ counts\,s$^{-1}$, and dead time occupies 69\% of 1-s exposure.
Since the data are highly affected by the pile-up effect, we developed a data-acquisition pile-up simulator of the WAM system to correct for the pile-up effect.
The detailed design of the simulator is reported in Appendixes \ref{sec:developingSim} and \ref{sec:demonstrationSim}.
The tool simulates the analog signal processing in the on-board electronics.
The input data are a background spectrum, a light curve, response matrix, and spectral 
model; and the output is a spectral model affected by the pile-up effect with instrumental response.
By comparing the output model that considers the pile-up effect with the observed real spectrum, we searched the best-fit parameters by using a Monte-Carlo approach and calculated the chi-squared values.

\subsection{Time-averaged Spectral Fitting}
\label{sec:withPileup}
We performed time-averaged spectral analysis using our data-acquisition pile-up simulator described in \S \ref{sec:respPileup}.
As the first step, we applied four single-component models consisting of a blackbody (BB), an OTTB, a power-law (PL), and a PL with an exponential cut-off (PLE).
None of these models yielded an acceptable fit, and gave chi-squared over degree-of-freedom values of $\chi^{2}/{\rm d.o.f} = 620.3/50$, $631.1/50$, $114.5/50$, and $128.3/49$, respectively.
Therefore, single-component models were not acceptable.
Next, we utilized two-component combination models consisting of 2BB, a BB plus an OTTB (BB$+$OTTB), a BB plus a PL (BB$+$PL), and a BB plus a PLE (BB$+$PLE).
The fitting results for the two-component models are shown in Table \ref{tab:fit}, which shows that BB$+$OTTB and BB$+$PLE models are reasonable.
The measured fluence of the two best fitting models in the range of 25\,keV to 2\,MeV, compared with the ACS measurements \citep{mereghetti2009a}, are  about $3.0 \times 10^{-4}$\,\ergcm\ and about $2.7\times 10^{-4}$\,\ergcm, respectively.
These values are consistent with the lower limit of $> 4.6 \times 10^{-5}$\,\ergcm\ provided in \citet{mereghetti2009a}.

Normally, a hard X-ray instrument with range 160\,keV--6.2\,MeV, such as the WAM detector, cannot constrain soft components such as those in the BB model with a temperature of 4.0--13\,keV.
However, in this observation of the hardest burst, the piled-upped events from the softer energy band below the lower threshold of the detector carry information about the soft component; that is, the WAM limits the parameters of the temperature of the BB component.
Note that the systematic error in determining the fitting model parameters by the pile-up simulator is not included in the above results. In principle, we cannot perform a systematic study of the pile-up simulator for this kind of very rare bright event.

\begin{table}
  \begin{center}
    \caption{Spectral parameters\footnotemark[$*$]}
    	\label{tab:fit}
    \begin{tabular}{lcccc}
      \hline\hline
      & \multicolumn{4}{c}{ Average }\\ \cline{2-5}
	Model & 2BB & BB$+$OTTB & BB$+$PL & BB$+$PLE \\
      \hline
	$kT_{\rm BB}$ (keV) 
		& $5.8 ^{+0.5}_{-0.1}$
		& $12.6 ^{+1.1}_{-1.5}$
		& $4.0 ^{+0.7}_{-0.5}$
		& $13.1 ^{+1.0}_{-0.7}$\\
	$R_{\rm BB}$ (km) \footnotemark[$\dagger$] 
		& $112 ^{+73}_{-68}$
		& $10 \pm6$
		& $289 ^{+183}_{-182}$
		& $9.1 ^{+5.6}_{-5.1}$\\
	$kT_{\rm BBhigh}$ (keV) 
		& $47.7 ^{+0.5}_{-1.7}$
		& $\cdots$ & $\cdots$ & $\cdots$ \\
	$R_{\rm BBhigh}$ (km) \footnotemark[$\dagger$] 
		& $0.30 \pm0.17$
		& $\cdots$ & $\cdots$ & $\cdots$\\
	$kT_{\rm OTTB}$ (keV) 
		& $\cdots$  
		& $405 ^{+72}_{-29}$
		& $\cdots$	& $\cdots$\\
	$\Gamma$ 
		& $\cdots$ & $\cdots$ 
		& $3.02^{+0.10}_{-0.03}$
		& $0.89 ^{+0.51}_{+0.24}$\\
	$E_{\rm cut}$ (keV) 
		& $\cdots$ 	& $\cdots$	& $\cdots$ 
		& $283 ^{+79}_{-29}$\\
	$F$ ($10^{-6}$\,\ergcms) \footnotemark[$\ddagger$] 
		& $18.2 \pm5.5$
		& $10.6 ^{+4.0}_{-3.4}$
		& $17.5 ^{+5.4}_{-5.7}$
		& $9.82 ^{+4.53}_{-4.21}$\\
	$\chi^{2}$/d.o.f \footnotemark[$\S$] 
		& 325.1/48
		& 74.5/48
		& 91.4/48
		& 73.0/47\\
      \hline
      
      & \multicolumn{4}{c}{ Peak }\\ \cline{2-5}
      Model & 2BB & BB$+$OTTB & BB$+$PL & BB$+$PLE\\
      \hline
	$kT_{\rm BB}$ (keV) 
	& $5.4 \pm0.1$
	& $18.9 ^{+3.8}_{-7.4}$
	& $6.8 \pm0.1$
	& $19.7 ^{+0.8}_{-7.4}$\\
	$R_{\rm BB}$ (km) \footnotemark[$\dagger$] 
	& $268 ^{+169}_{-151}$
	& $3.4 ^{+2.9}_{-2.1}$
	& $97 ^{+54}_{-56}$
	& $3.3 ^{+2.9}_{-2.2}$\\
	$kT_{\rm BBhigh}$ (keV)  
	& $106 ^{+2}_{-4}$
	&$\cdots$&$\cdots$&$\cdots$\\
	$R_{\rm BBhigh}$ (km) \footnotemark[$\dagger$]  
	& $0.09 ^{+0.05}_{-0.06}$
	&$\cdots$&$\cdots$&$\cdots$\\
	$kT_{\rm OTTB}$ (keV)  
	&$\cdots$
	& $346 ^{+36}_{-27}$
	&$\cdots$&$\cdots$\\
	$\Gamma$  
	& $\cdots$ 	& $\cdots$
	& $2.36 ^{+0.06}_{-0.02}$
	& $0.86 ^{+0.38}_{-0.36}$\\
	$E_{\rm cut}$ (keV)  
	& $\cdots$ 	& $\cdots$ 	& $\cdots$
	& $270 ^{+88}_{-65}$\\
	$F$ ($10^{-6}$\,\ergcms) \footnotemark[$\ddagger$]  
	& $65.8 ^{+20.3}_{-19.8}$
	& $24.1 ^{+13.5}_{-10.3}$
	& $49.5 ^{+14.9}_{-15.4}$
	& $24.3 ^{+23.6}_{-14.9}$\\
	$\chi^{2}$/d.o.f \footnotemark[$\S$]  
	& 234.6/46
	& 79.2/46
	& 291.2/46
	& 83.9/45\\
      \hline
      
      & \multicolumn{4}{c}{ Tail }\\  \cline{2-5}
      Model & 2BB & BB$+$OTTB & BB$+$PL & BB$+$PLE\\
      \hline
	$kT_{\rm BB}$ (keV) 
		& $9.9 ^{+0.5}_{-0.3}$ 
		& $9.0 ^{+1.2}_{-4.6}$
		& $15.2 ^{+11.4}_{-6.0}$
		& $15.2 ^{+0.3}_{-4.3}$\\
	$R_{\rm BB}$ (km) \footnotemark[$\dagger$] 
		& $5.3 \pm3.0$ 
		& $7.8 ^{+6.7}_{-5.0}$
		& $2.4 ^{+1.6}_{-2.1}$
		& $2.7 ^{+2.2}_{-1.8}$\\ 
	$kT_{\rm BBhigh}$ (keV)  
                & $39.0 ^{+3.2}_{-0.6}$
                & $\cdots$ & $\cdots$ & $\cdots$\\
	$R_{\rm BBhigh}$ (km) \footnotemark[$\dagger$]  
                & $0.19 \pm0.11$
                & $\cdots$ & $\cdots$ & $\cdots$\\
	$kT_{\rm OTTB}$ (keV)  
		& $\cdots$ 
		& $81.2 ^{+11.9}_{-10.0}$
		& $\cdots$ & $\cdots$\\
	$\Gamma$  
		& $\cdots$ 
		& $\cdots$
		& $3.63 ^{+0.16}_{-0.45}$
		& $1.48 ^{+0.37}_{-0.25}$\\
	$E_{\rm cut}$ (keV)  
		& $\cdots$ & $\cdots$ & $\cdots$ 
		& $120 ^{+10}_{-43}$\\
	$F$ ($10^{-6}$\,\ergcms) \footnotemark[$\ddagger$]  
		& $2.62 ^{+0.81}_{-0.88}$
		& $2.65 ^{+1.44}_{-0.93}$
		& $2.26 ^{+5.37}_{-1.53}$
		& $2.41 ^{+2.01}_{-1.71}$\\
	$\chi^{2}$/d.o.f \footnotemark[$\S$]  
		& 22.4/23
		& 19.9/23
		& 28.9/23
		& 21.2/22\\
      \hline
    \end{tabular}
  \end{center}
\footnotemark[$*$]: BB, OTTB, PL, and PLE represent blackbody, optically thin thermal bremsstrahlung, power-law, and power-law with exponential cut-off, respectively.\\
\footnotemark[$\dagger$]: Blackbody radius assuming a distance of 4 kpc. Uncertainty of absolute flux in the response matrices of 30\% is included.\\
\footnotemark[$\ddagger$]: Soft gamma-ray flux in the range of 160\,keV--6.2\,MeV. Uncertainty of absolute flux in the response matrices of 30\% is included.\\
\footnotemark[$\S$]: Degrees of freedom.\\
\end{table}

\begin{figure}
  \begin{center}
    \begin{center}
	\FigureFile(80mm,80mm){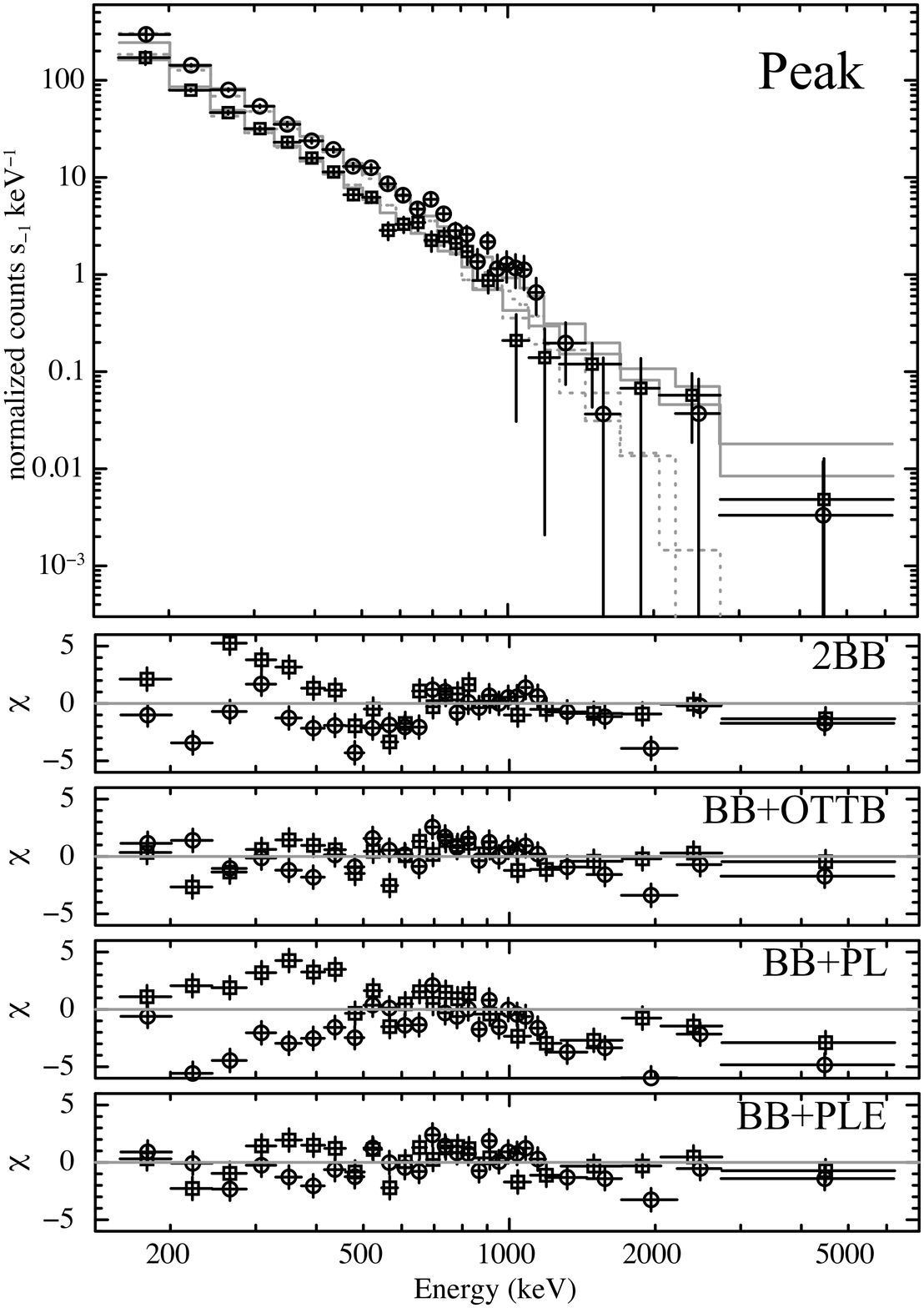}
	\FigureFile(80mm,80m){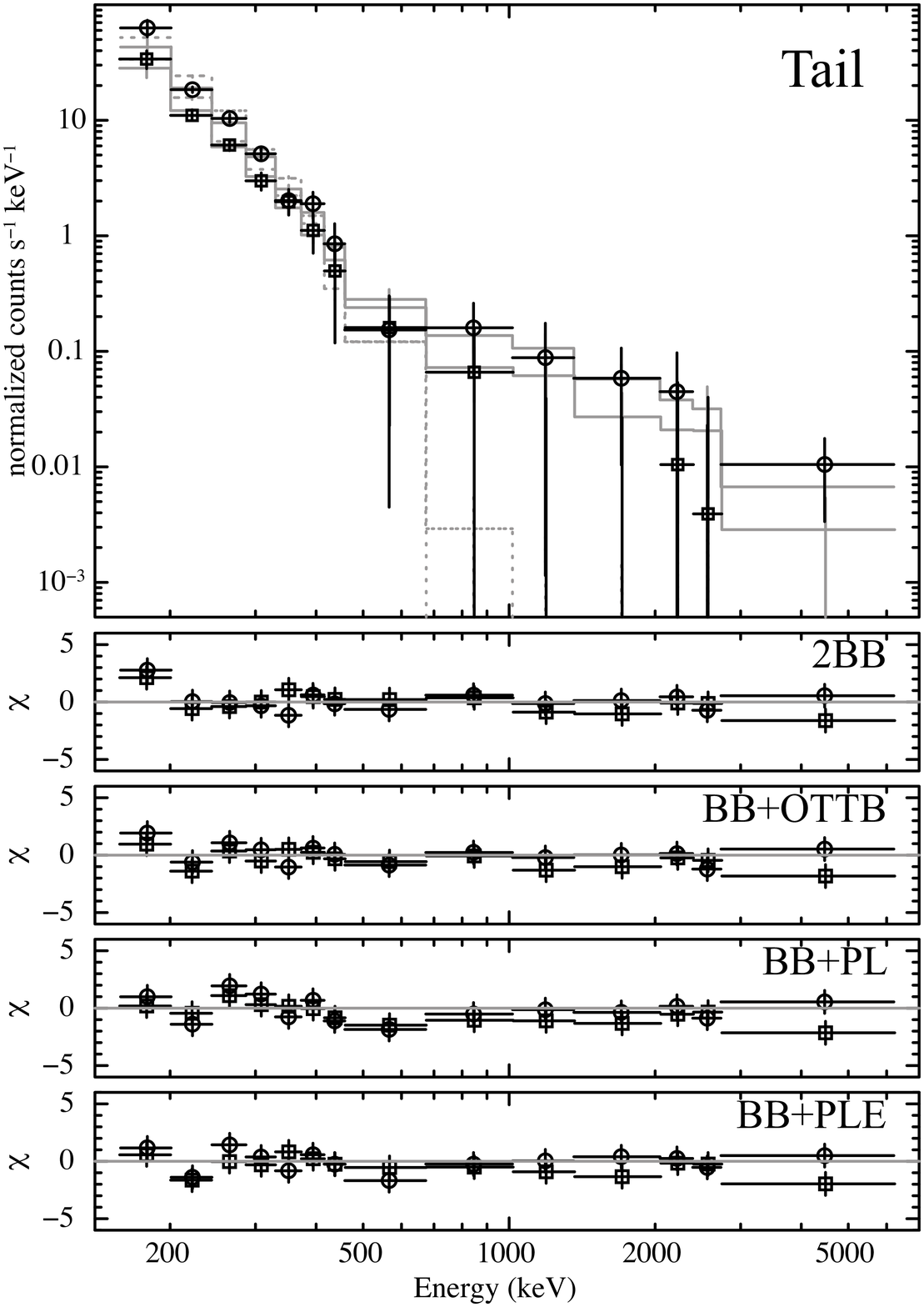}
    \end{center}
    \caption{Time-resolved spectra of the hardest short burst of WAM-0 (circles) and WAM-1 (squares) of peak (left) and tail (right) intervals.
    Upper and lower panels represent background-subtracted spectra using the BB$+$OTTB model and the residuals from best-fit models affected by the pile-up effect, respectively.
    Models before and after being affected by the pile-up effect are shown as dotted and solid lines, respectively.}
    \label{fig:bestFit}
  \end{center}
\end{figure}

\subsection{Time-resolved Spectral Fitting}
\label{sec:timeResoSpec}

To investigate temporal variations in the spectral parameters, we divided the hardest burst into two regions and performed spectral fitting using the eight models.
Extraction of the spectral information was performed following the same procedure as for the averaged spectra (in \S\ref{sec:dataSelection} and \S\ref{sec:respPileup}).
Spectra of the peak and the tail region are better fitted by two-component models than single-component models, as already reported for the time-averaged spectra (\S \ref{sec:withPileup}).
Table \ref{tab:fit} and Figure \ref{fig:bestFit} summarize the fitting results for the two-component models.
The best-fit models in both time intervals, BB$+$OTTB and BB$+$PLE, yield similar parameters for BB temperature and peak energy $kT_{\rm OTTB}$/$E_{\rm cut}$, and the spectral shapes of the PLE component for the photon index $\Gamma = 0.86 ^{+0.38}_{-0.36}$ and $1.48 ^{+0.37}_{-0.25}$ are approximately equal to the OTTB shapes.
From the above results, we determined that the best-fit models are the BB$+$OTTB and BB$+$PLE models.
A detailed discussion is given in \S\ref{sec:discussion}.

\section{Discussion}
\label{sec:discussion}
\noindent
\subsection{Hard X-ray Spectral Shape}
\label{discussion1}

\begin{figure}
  \begin{center}
    \begin{center}
	\FigureFile(80mm,80mm){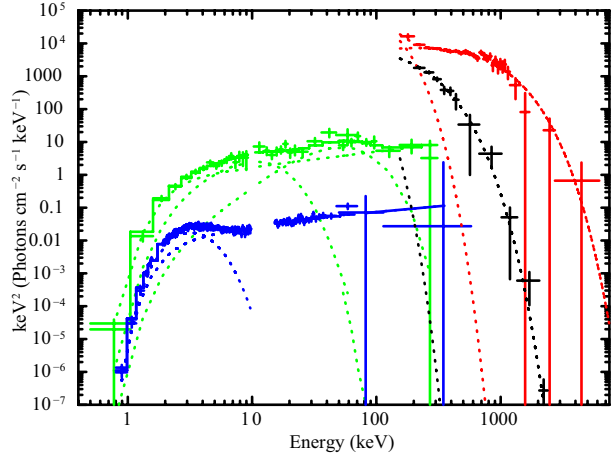}
    \end{center}
    \caption{$\nu F_{\nu}$ spectra of Suzaku view during 2009 activity of \axp. Blue and green represent spectra of persistent emission and accumulated weak short bursts, respectively, from \citet{enoto2010a} and \citet{enoto2012}. Red and black represent the WAM-0 spectra of the peak and the tail region, as fitted by the BB$+$OTTB model. The systematic error is not included in the WAM spectra.}
    \label{fig:suzakuView}
  \end{center}
\end{figure}

We presented a spectral analysis of the hardest burst, which occurred at UTC 2009 January 22 06:45:13, using the WAM spectra with the newly developed pile-up simulator (\S \ref{sec:respPileup}).
In this work, thanks to the ability of the WAM to perform wide-band spectroscopy over a large effective area, we succeeded in revealing the spectral shapes of the hardest burst in the sub-MeV (160\,keV--6.2\,MeV) band, and found that the spectra were well reproduced by the two-component combinations of the BB$+$OTTB and BB$+$PLE models (\S\ref{sec:timeResoSpec}).
As shown in \S\ref{sec:timeResoSpec}, the model parameters, such as the OTTB temperatures $kT_{\rm OTTB}$, and the cut-off energies $E_{\rm cut}$, dramatically changed during the burst in 2 s.
Figure \ref{fig:suzakuView} summarizes the $\nu F_{\nu}$ spectra from the BB$+$OTTB model during the peak and tail epochs with the WAM for comparison with during the persistent emission \citep{enoto2010a} and the accumulated weak short bursts obtained by the follow-up Suzaku observation performed about one week after the hardest burst \citep{enoto2012}.

\begin{figure}
  \begin{center}
   \begin{center}
	\FigureFile(80mm,80mm){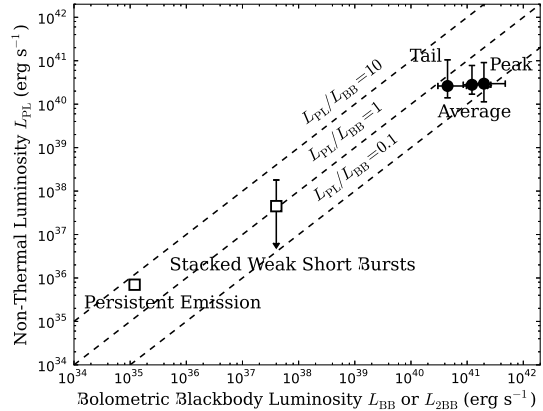}
   \end{center}
    \caption{Relation between the bolometric BB luminosity and that of the hard X-ray PL component in the 1--300\,keV energy range.
	Points of persistent emission and stacked weak short bursts (squares) are taken from Figure 15 of \citet{enoto2012}.
	Our results for the BB$+$PLE model (circles) are plotted with the luminosities including the systematic error of absolute flux of 30\%.}
     \label{fig:luminosityRelation}
 \end{center}
\end{figure}

In the BB component, the temperatures of the BB$+$OTTB and BB$+$PLE models ($kT_{\rm BB} \sim$ 9--20\,keV; Table \ref{tab:fit}) are comparable with the average temperatures from the 2BB model with the Fermi GBM ($\sim 5$\,keV and $\sim 14$\,keV; \cite{horst2012}) and consistent with that for the accumulated weak short bursts ($\sim 13$\,keV; \cite{enoto2012}).
Therefore, although the temperature may not change among these various fluxes by a difference of two orders of magnitude, this flux dependency is not obvious because the values from the WAM should have some systematic uncertainties from the pile-up simulator.
Similarly, for the hard component of the BB$+$PLE model during the peak and tail intervals, the photon index ($\Gamma \sim $0.9--1.5) and cut-off energy ($E_{\rm cut} \sim 283$\,keV) are consistent with those in the persistent emission ($\Gamma \sim $1.3--1.5 and $E_{\rm cut} > 200$\,keV; \cite{enoto2010a}).
Therefore, the spectral shapes of each component are seen to be stable among different phases.
Furthermore, in order to compare the energy partition rate to each radiation component among the three different emissions from the object, Figure \ref{fig:luminosityRelation} shows a scatter plot of the luminosity in the bolometric BB luminosity ($L_{\rm BB}$ or $L_{\rm 2BB}$) and the 1--300\,keV luminosity ($L_{\rm PL}$) of the PL component.
The data for our work on the hardest burst are calculated by extrapolating the energy range down to 1\,keV by using the best-fit parameters of the BB$+$PLE model (Table \ref{tab:fit}).
As a result, the empirical correlation for the luminosities between BB and non-thermal components can be extended into the four-orders-of-magnitude brighter region, although the similarities on the spectral shape below about $10^{38}$\,\ergs\ are already indicated in \citet{enoto2012}.
Our results provide a second example of a magnetar whose spectral shape in the hard X-ray to soft gamma-ray band is stable among various luminosity ranges, after the first report on SGR~0501$+$4516 \citep{nakagawa2011}.
These results further support the common radiation mechanism in observationally different emissions.

\subsection{Fluence for the Light Echoes}
\label{discussion2}

Although \citet{tiengo2010} suggested that the hardest burst is one candidate for X-ray scattering echoes, the exact fluence has not been measured.
According to this suggestion, unabsorbed total energy between $10^{44}$\,erg and $2 \times 10^{45}$\,erg is necessary to generate echoes.
However, our measurements using the BB$+$OTTB and BB$+$PLE models suggest a total energy of $(6.8 ^{+2.7}_{-2.3}) \times 10^{41}\,(\frac{d}{4 {\rm~kpc}})^{2}$\,erg and $(6.0 ^{+2.1}_{-1.9}) \times 10^{41}\, (\frac{d}{4 {\rm~kpc}})^{2}$\,erg from 1\,keV to 10000\,keV range, respectively, assuming isotropic emissions.
Although systematic error from the pile-up simulator is not included in the energies, we succeeded in measuring the emission energy accurately through pile-up correction and found that none of the models can satisfy the conditions for forming echoes.
More precisely, our measurement is performed in the limited-energy bandpass (160\,keV--6.2\,MeV), and the total fluence could become larger if there were to be another emission component below 160\,keV; however, it is not feasible to enhance the fluence by two or three orders of magnitude.

Another candidate other than the hardest burst could be burst No. 149, defined in \citet{mereghetti2009a}, which has the highest fluence among bursts detected by the ACS and occurred at UTC 06:48:04 after about 170 s from the hardest burst followed by a long pulsating tail ($\sim{}8$ s).
The emission energy is already estimated as about $2.4 \times 10^{43} ( \frac{d}{10 {\rm~kpc}})^{2}$ erg ($\sim 3.8 \times 10^{42} ( \frac{d}{4 {\rm~kpc}})^{2}$ erg) in \citet{mereghetti2009a}, and this energy is likewise not sufficient for generating light echoes.
The light echoes are attributed not to the recorded burst tail but to an unrecorded initial spike which caused the safety switch off of the WAM and saturated the ACS.
Therefore, to reveal the origin of the light echoes, the unmeasured soft X-ray spectral shape, especially in the initial spike, is required.

\subsection{Time Evolution}
\label{discussion3}

The detailed light curve of the hardest burst observed by the ACS (Figure \ref{fig:detailedLightCurve}) exhibits several time-variable components: a bright initial spike; a slow decay; and a rapid disappearance.
Actually, the time-resolved spectra observed by the WAM (\S \ref{sec:timeResoSpec}) also exhibited extreme changes in spectral parameters such as OTTB temperature, and the cut-off energy of PLE component.
Such temporal and spectral variations are typical characteristics of giant flares, but no such variations are observed in the shorter time scale (ex., 1.5\,s; \cite{mereghetti2009a}) than the spin period of the object or dimmer radiation energy than typical flux of giant flares (10$^{44}$\,erg).
The properties of the hardest burst are as if the burst was a ``small-scale giant flare'', and match those of the intermediate flares \citep{woods2006}.
Therefore, this hardest burst could be the missing link between giant flares and intermediate flares, which are expected to have the same radiation mechanisms because the former commonly follow the latter by several days or months. However, a giant flare has not been observed from \axp\ to date.

We attempt to explain the ACS light curve (Figure \ref{fig:detailedLightCurve}) by using the same interpretation of giant flares as \citet{thompson2001}, which describes the rapid phenomena as a cooling of a trapped pair-photon fireball forced in a strong magnetic field: $L_{\rm X}(t) = L_{\rm X}(0) (1-t/t_{\rm eval})^{\chi}$, where $t_{\rm eval}$ and $\chi$ indicate the evaporation time and benchmark of the fireball geometry, respectively.
According to \citet{thompson2001}, the index $\chi$ indicates the dimension of the fireball surface: $\chi = 2$ for spherical, $\chi = 1$ for cylindrical, and $\chi = 0$ for thin slab.
The trapped fireball model is nicely fitted to the giant flares \citep{feroci2001,hurley2005} and some intermediate flares \citep{olive2004}.
In typical short bursts, an exponential function is often utilized to represent the decay shape.
In our analysis, the ACS light curve of the hardest burst is found to be well described by a combination of the above two components.
The best-fit model for the time interval of UTC 06:45:14.1--16.2, excepting the initial bright spike, is shown in Figure \ref{fig:detailedLightCurve}.
The fitting yields the following parameters: a time constant $\tau_{\rm exp} = -0.38 \pm 0.02$ s of the exponential component, an evaporation time $t_{\rm eval} = 1.35 \pm 0.04$ s and an index $\chi = 0.24 \pm 0.06 $ of the trapped fireball component with degree of freedom over chi-square of 104.4/42.

Comparing the light curve fitting with the spectral fitting results (\S \ref{sec:timeResoSpec}), we can consider that the trapped fireball and the exponential components reflect temporal flux variation of the BB and the OTTB/PLE component, respectively.
In the BB$+$PLE spectral fitting during the tail interval, the calculated photon numbers of the BB and the PLE component are about $36$ and $17$, respectively, extrapolating the spectral model to the ACS energy range (80\,keV--8\,MeV).
The ratio of photon numbers is qualitatively consistent, with the trapped fireball component having a higher count rate than the exponential component during the tail interval in the light curve fitting (Figure \ref{fig:detailedLightCurve}).

Previous reported indices $\chi$ of giant flares of SGR~1806$-$20 on 2004 December 27 and SGR~1900$+$14 on 1998 August 27 are 1.5 \citep{hurley2005} and 3 \citep{feroci2001}, and a intermediate flare from SGR~1900$+$14 is described by two trapped fireball of $\chi = 0.4$ and 0.1 \citep{olive2004}.
Therefore, our results for the hardest burst of \axp\ are similar to intermediate flares.
Among these little samples, we found that the index $\chi$ of the intermediate flares tends to be smaller than giant flares.
Since the duration of the ACS light curve of 1.45 s \citep{mereghetti2009a} is significantly longer than the half rotation period (2.1/2 $=$ 1.05 s), we should note that the time-variation may not reflect intrinsic evolution of the emission.\\[3ex]%

This work was supported in part by Grant-in-Aid for the Japan Society for the Promotion of Science (JSPS) Fellows (No. 24-10233, T. Y.), Grants-in-Aid for Scientific Research (B) from the Ministry of Education, Culture, Sports, Science, and Technology (MEXT) of Japan (No. 23340055, Y. T.; No. 22340039, M. S. T.), a Grant-in-Aid for Young Scientists (A) from MEXT (No. 22684012, A. B.), and JSPS KAKENHI (No. 24540309, Y. E. N.).

\appendix

\section{Development of Pile-up Simulator and Estimation of Influence of The Effect}
\label{sec:developingSim}
If more than one photon comes into the detectors in a time interval shorter than the processing pitch of the on-board electronics, these events are not able to be separated into individual events and thus become treated as a high-energy event or a non-X-ray event.
In other words, the spectral shape becomes harder in the case of very bright sources, and thus estimation of the pile-up effect on spectra is important.
We therefore developed a pile-up simulation code based on the C++ programming language to reproduce the on-board analog electronics (TPU: Transient data Processing Unit, \cite{takahashi2007}) and to correct the effect.
We utilized the SLLIB and SFITSIO\footnote{http://www.ir.isas.jaxa.jp/\~{}cyamauch/sli/index.html} libraries for reading and writing the Flexible Imaging Transport System (FITS) data format files.
The inputs are a background spectrum, a light curve, a response matrix, and the spectral model, and the output is the spectral model affected by the pile-up effect.

The simulator plays the following role.
First, it reads the input files and calculates the incident event number.
Second, based on the shapes of the input spectra and the light curve, the simulator randomizes the arrival time and photon energy.
The time and energy information are applied to individual incident events.
Third, these events pass thorough the reproduced processing algorithm of the analog electronics TPU.
It is at this point that they are affected by the pile-up effect.
Moreover, the simulator accumulates the spectrum from those events.
Finally, these three steps are repeated 128 times with different random seeds.
The simulator calculates an average spectrum assuming a Poisson distribution, which it writes out to a FITS format file as the output model.

\begin{figure}
  \begin{center}  
  	\begin{center}
	\FigureFile(80mm,80mm){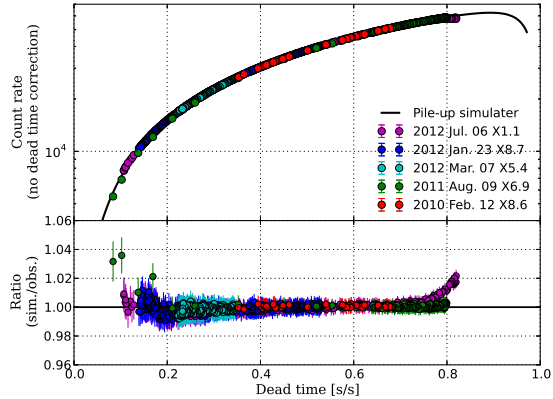}
	\end{center}
    \caption{Relationship between dead time and count rate, and comparison of results of the pile-up simulator with observed real data of bright solar flares observed by WAM-0 after 2009.
    	(Top) Red, green, blue, cyan and magenta circles are data for solar flares on 2010 February 12, 2011 August 09, 2012 January 23, 2012 March 07 and 2012 July 06, respectively.
	Solid black line is the result of the pile-up simulator.
	(Bottom) Ratio of count rate between the observed and simulated data.
	We found that the reproducibility of the pile-up simulator is much better than the uncertainty of the response matrixes of 30\%.}
    \label{fig:livetime}
  \end{center}
\end{figure}

The absolute flux is estimated by the count rate and the dead time reported by the on-board electronics TPU. Similarly, the pile-up simulator also reports both the count rate and dead time as a response to the incident count rate, which is unknown. 
In order to verify the pile-up simulator, we checked if the relation between the count rate observed and dead time from the simulator matched the relationship by TPU even for brighter than usual cases. For this, we used five bright solar flares observed by the WAM after 2009 as shown in Figure \ref{fig:livetime}.
The dead time and count rate values were extracted from public TRN data of WAM-0 with temporal resolution of 1\,s.
These solar flares were observed in 2010 February 12, 2011 August 09, 2012 January 23, 2012 March 07 and 2012 July 06, and are classed as Goes class M8.3, X6.9, M8.7, X5.4 and X1.1, respectively.
As shown in Figure \ref{fig:livetime}, the pile-up simulator well reproduced the real data, and the reproducibility is a lot lower than the uncertainty of the response matrixes of 30\%.
In particular, the dead times of the hardest burst spectra for 1.21\,s (WAM-0) and 1.14\,s (WAM-1) to 2 \,s exposures correspond to 0.61\,s\,s$^{-1}$ and 0.57\,s\,s$^{-1}$, respectively, and in Figure \ref{fig:livetime} reproducibility around these dead times is within 1\%.

\section{Demonstration of Pile-up Simulator}
\label{sec:demonstrationSim}
\begin{figure}
  \begin{center}  
  	\begin{center}
	\FigureFile(80mm,80mm){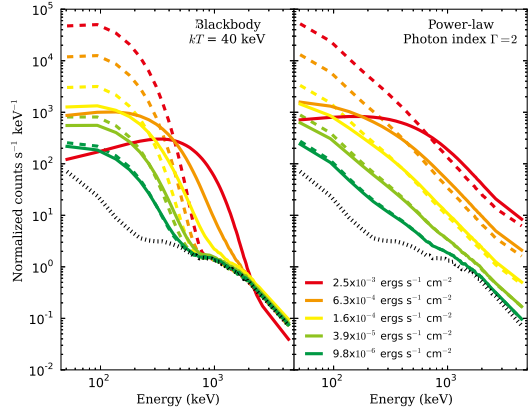}
	  \end{center}
    \caption{Demonstration of the pile-up simulator.
	Left and right panels show estimated observed spectra affected by the pile-up effect (solid lines) and original spectra unaffected by the effect (dashed lines) of BB of 40\,keV and PL function of photon index $\Gamma = 2.0$.
	These spectra have the same BB temperature or photon index, and only the source fluxes are different.
	Colors indicate different fluxes in the 50\,keV--5\,MeV energy range, with green to red corresponding to from $9.8 \times 10^{-6}$ to $2.5 \times 10^{-3}$\,\ergcms, respectively.
	Dotted lines are background level.}
    \label{fig:demonstration}
  \end{center}
\end{figure}

The weight of inflection of the spectral shape due to the pile-up effect depends on the light curve, response matrix, background spectrum, source spectrum, and brightness.
We demonstrated the dependence on brightness as shown in Figure \ref{fig:demonstration} using the pile-up simulator with two spectral models of BB of 40\,keV and single PL function of photon index $\Gamma=2.0$ and the data of the hardest burst; the light curve of the SPI-ACS, the calculated response matrix, the estimated background spectrum of WAM-0 in \S \ref{sec:dataSelection}.
These demonstrations indicate that observed spectral shapes are greatly changed by the pile-up effect depending on the input photon number.
In particular in the case of higher flux (orange and red lines in Figure \ref{fig:demonstration}), the flux in the lower energy range is strongly decreased and the higher energy range is increased compared to the original spectrum unaffected by the pile-up effect.
Observed count rate of the hardest burst is comparable with the lines of second higher flux cases (orange lines in Figure \ref{fig:demonstration}).
Therefore to investigate original spectral shape and flux of the hardest burst, correction of the pile-up effects is important.

\end{document}